 \documentclass[aoas,preprint]{imsart}
\usepackage{amsthm,amsmath,chicago}
\RequirePackage{graphicx,color,verbatim,multirow,pgf,pgfarrows,pgfnodes}

\startlocaldefs
\endlocaldefs

\begin{document}

\begin{frontmatter}

\title{A Hierarchical Model for Estimating HIV Epidemics}
\runtitle{A Hierarchical Model for Estimating HIV Epidemics}


\author{\fnms{Le} \snm{Bao$^1$}\ead[label=e1]{lebao@psu.edu}}
\and
\author{\fnms{Mary} \snm{Mahy$^2$}}
\and
\author{\fnms{Xiaoyue} \snm{Niu$^1$}}
\and
\author{\fnms{Tim} \snm{Brown$^3$}}
\and
\author{\fnms{Peter D.} \snm{Ghys$^2$}}
\affiliation{Department of Statistics, Penn State University, University Park, PA, USA$^1$\\
UNAIDS, Strategic Information and Evaluation Department, Geneva, Switzerland$^2$\\
Population and Health Studies, East-West Center, Honolulu, HI, USA$^3$\\
\printead{e1}}

\runauthor{L. Bao et. al.}

\begin{abstract}
As the global HIV pandemic enters its fourth decade, increasing numbers of surveillance sites have been established which allows countries to look into the epidemics at a finer scale, e.g. at sub-national level. However, the epidemic models have been applied independently to the sub-national areas within countries. An important technical barrier is that the availability and quality of the data vary widely from area to area, and many areas lack data for deriving stable and reliable results. To improve the accuracy of the results in areas with little data, we propose a hierarchical model that utilizes information efficiently by assuming similar characteristics of the epidemics across areas  within one country. The joint distribution of the parameters in the hierarchical model can be approximated directly from the results of independent fits without needing to the refit the data. As a result, the hierarchical model has better predictive ability than the independent model as shown in examples of multiple countries in Sub-Saharan Africa.
\end{abstract}


\end{frontmatter}

\section{Introduction}
\label{sect-Introduction}
Since the first case reported in 1981 \cite{CDC1981}, the global HIV epidemic has become one of the greatest threats to human health and development. The number of people living with HIV worldwide continued to grow, reaching 35 million in 2013, about three times more than in 1990. There have been 78 million infected with HIV and over 29 million AIDS-related deaths so far. The response to the HIV epidemic has been mixed, with progress being made to reduce new infection and rapid improvements in survival rates among AIDS patients as antiretroviral therapy has become available in recent years \cite{UNAIDS2014}. 

Countries need to ground their HIV strategies in an understanding of their own epidemics and their national responses. Reliable estimation and prediction of the HIV epidemic can help policy makers and program planners efficiently allocate their resources, plan and manage the intervention, treatment and care programs, evaluate their effort, and raise funds. Therefore, accurate estimation and projection of the epidemic is the foundation of all HIV-related studies \cite{WHO2013}. In addition, countries -- especially large countries and countries with geographically heterogeneous epidemics -- need to understand the variation of the impact within the country by geographic area and population \shortcite{Mahy2014}. Having surveillance data and HIV estimates available by sub-national entities helps district-level managers focus their limited resources \cite{UNAIDS2013}. 


Most countries (158 countries submitted files in 2014) use Spectrum software to estimate the impact of HIV on their population at the national level \shortcite{Case2014}.  Spectrum relies on data from HIV surveillance systems and the Estimation and Projection Package (EPP) to estimate trends in HIV incidence over time \shortcite{Spectrum2014,Brown2014,Stover2014}. The UNAIDS Reference Group on Estimates, Modelling and Projections has developed, and continues to refine Spectrum/EPP for estimation and short-term prediction of HIV trends \shortcite{Ghys2004,Brown2006,Brown2010,Brown2014}. Due to the paucity of reliable information on the incidence of HIV in most countries, sentinel surveillance systems for HIV were designed to provide information on prevalence trends. EPP estimates trends in HIV prevalence, incidence and mortality from these surveillance data from 1970 through the current year, and makes short term projections.  

As the global HIV pandemic enters its fourth decade, increasing numbers of surveillance sites have been established which allows countries to look into the epidemics at a finer scale, e.g. at sub-national level.   However, the estimation and projection process has been implemented independently for areas in those countries. The quality of HIV surveillance data used for these models is variable \shortcite{Calleja2010}.  Quality has multiple dimensions including the number of years of data to show trends over time, the representativeness of the data across the country and districts and the accuracy of those data \shortcite{Lyerla2008}. When countries/districts do not have high quality data the models produce inaccurate results with large uncertainty bounds. Most countries with generalized epidemics (epidemics in which HIV transmission is primarily in the general population through heterosexual sex) have historical HIV surveillance data from women attending antenatal clinics from year 2000 to now.  In addition, all but four generalized epidemic countries have household surveys that measure the HIV prevalence in the national population as well as sub-national estimates to the first sub-national administrative level \shortcite{Marsh2014}.



 
One way to improve the accuracy of the estimates is to borrow information from neighboring administrative areas in the case of sub-national estimates. For example many areas have few data points early in the HIV epidemic (between 1980 and 2000).  In such settings, if the data from neighboring areas is assumed to have similar trends, the trends from those areas can be used to inform the trends for the same time period in the data-free area. In this paper we describe a method to use a hierarchical model to share information across datasets. In the proposed model, the epidemiologically implausible trajectories fit in the area with little data would be corrected by the area with rich data. The difficulty of utilizing the hierarchical model is that it will not be a standard mixed effects model since the underlying epidemic model is described by differential equations, and the computational cost of fitting such a hierarchical model is high because the model parameters do not have analytic solutions. In addition, national officials are also interested in comparing the results between the independent fits and the hierarchical model fits, as well as exploring different settings in the hierarchical model. In this article, we propose an importance sampling method that draws posterior samples of parameters in the hierarchical model without needing to refit the data. 

In Section 2 we describe the EPP model used by UNAIDS, the hierarchical model, the parameter estimation, and the evaluation procedure in detail. In Section 3, we give results for 20 countries in Sub-Saharan Africa, and in Section 4 we offer conclusions and discussion for future work. 



\section{Approach}
\label{sect-Approach}
In this section, we first describe a dynamic model used by EPP for estimating and projecting the HIV epidemic from prevalence data. We then introduce the hierarchical model that allows information to be shared across areas with similar epidemiological patterns. After that we propose a novel parameter estimation method for the hierarchical model that avoids additional runs of the dynamic model. Finally we discuss the model evaluation procedure.

\subsection{The EPP Model}
The Estimation and Projection Package (EPP) is based on a simple susceptible-infected-removed (SIR) epidemiological model. The population being modeled is aged between 15 and 49, and the population at time $t$ is divided into two groups: $Z(t)$ is the number of uninfected individuals, $Y(t)$ is the number of infected individuals. For parsimony in presentation, we describe here a simplified version of the model without the details on CD4 progression that are included in the actual EPP model. The simplified dynamics are as follows:
\begin{equation}
\left\{\begin{array}{ccc}
\frac{dZ(t)}{dt} & = & E(t) - \frac{r(t) Y(t) Z(t)}{N(t)} - \mu(t) Z(t) - \frac{a_{50}(t)Z(t)}{N(t)} +  \frac{M(t)Z(t)}{N(t)}, \\
\frac{dY(t)}{dt} & = & \frac{r(t) Y(t) Z(t)}{N(t)} - \textup{HIVdeath}(t) - \frac{a_{50}(t)Y(t)}{N(t)} +  \frac{M(t)Y(t)}{N(t)}, \\
\end{array}\right.
\end{equation}
where $N(t) = Z(t) + Y(t)$ is the total adult population size, $E(t)$ is the number of new adults entering the population, $\textup{HIVdeath}(t)$ is the number of HIV related deaths. There is a set of parameters defined by external life-tables such as $\mu(t)$, the non-AIDS mortality rate; $a_{50}(t)$, the rate at which adults exit the model after attaining age 50, and $M(t)$, the rate of net migration into the population. 

The infection rate, $r(t)$, is the expected number of persons infected by one HIV positive person in year $t$ in a wholly susceptible population $Z(t)$; and the parameters that drive the change of $r(t)$ are estimated within the EPP model. In the EPP 2011 model formulation, $r(t)$ is modeled as a random walk on the log scale: $\log r(t) - \log r(t-1) \sim N(0, \kappa^2)$ \shortcite{Bao2010sti,Bao2010rflex}. The r-trend model implemented in EPP 2013 introduces a systematic mean structure to the random walk model \cite{Bao2012rtrend} and assumes the yearly change of infection rate in Equation (2) is related with three components: the infection rate $r(t)$, the prevalence rate $\rho(t)= \frac{Y(t)}{N(t)}$, and a tendency for $ r(t)$ to reach a steady state $t_1$ years after the starting year of the epidemic $t_0$ because we have observed the prevalence stabilized in the later period of the epidemic for many countries.
\begin{equation}
\log r(t+1)-\log r(t) = \beta_1 \times (\beta_0 - r(t)) - \beta_2 \rho(t) + \beta_3 \gamma(t),
\end{equation}
where $\gamma_t = \frac{(\rho(t+1)-\rho(t)) (t-t_1)^+}{\rho(t)}$ implies the tendency of stabilization. Therefore, the r-trend model imposes some common structure on $r(t)$ across countries: $r_t$ declines if $r_t>\beta_0$, or $\rho_t$ is too high, or the relative increase of prevalence is too large. 

Moreover, ANC prevalence data tends to be biased upwards because the pregnant women are more sexually active. We let $\beta_4$ be the bias of ANC data with respect to prevalence data from  national population-based household surveys (NPBS) on the probit scale. The following informative prior distributions have been obtained from previously applying the model to epidemic datasets in a large number of countries \cite{Bao2012rtrend}.
Once given a set of input parameters, $\theta = (t_0, t_1, \beta_0, \beta_1, \beta_2, \beta_3, \beta_4)$, the dynamic model produces time courses of HIV prevalence, incidence, and mortality as the outputs. We then approximate the likelihood by modeling the prevalence on the probit scale and using a hierarchical model with random effects to take account of serial prevalence data collected at the same clinic or from the same type of survey \shortcite{Alkema2007}.

\subsection{The Hierarchical Model}
A hierarchical model can be useful in associating $\theta's$ among areas with similar epidemiological patterns, e.g. urban and rural areas in the same country. Let $\theta_{ij}$ be the $j$th local level parameter of area $i$, such that $\theta_{i.} = (t_{i0}, t_{i1}, \beta_{i0}, \beta_{i1}, \beta_{i2}, \beta_{i3}, \beta_{i4})$. To infer the relationships between $\theta_{i.}$ and $\theta_{i'.}$, we assume the following distributions for the hierarchical model:
\begin{equation}
\left\{\begin{array}{ccc}
\theta_{i.}|\mu & \sim & \textup{Norm}[\mu, \Sigma_1]\\
\mu & \sim & \textup{Norm}[\mu_0, \Sigma_0]
\end{array}\right.
\end{equation}
where $(\mu, \Sigma_1)$ are the mean and covariance of the local level parameter, $(\mu_0, \Sigma_0)$ are the prior mean and covariance for $\mu$. In the application to EPP model, we assume that $\Sigma_0$ and $\Sigma_1$ are diagonal matrices with diagonal elements $\{\sigma_{0j}^2\}$ and $\{\sigma_{1j}^2\}$. Figure 1 shows an example of parameter relationships in the hierarchical model, where $\theta_{1.}=\theta_{\textrm{urban}}$, $\theta_{2.}=\theta_{\textrm{rural}}$, $\mu=\theta_{\textrm{country}}$. We see that the sub-national parameter estimates affect each other through the national parameter estimates.

\begin{figure}[!h]
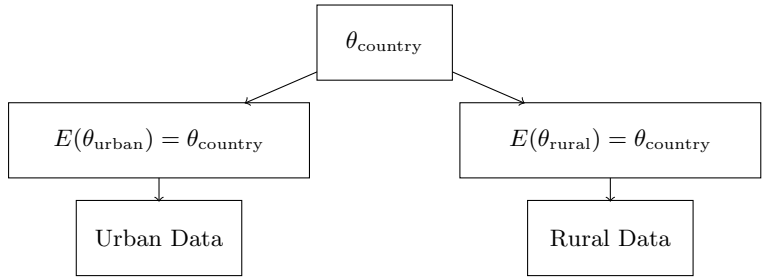

\vspace{7mm}
\begin{center}
\begin{pgfpicture}{-2.5cm}{-1.5cm}{2.5cm}{0.5cm}
\pgfnoderect{country}[stroke]{\pgfxy(0,0)}{\pgfpoint{1.8cm}{1cm}}
\pgfnoderect{urban}[stroke]{\pgfxy(-3,-1.3)}{\pgfpoint{4cm}{1cm}}
\pgfnoderect{rural}[stroke]{\pgfxy(3,-1.3)}{\pgfpoint{4cm}{1cm}}
\pgfnoderect{data1}[stroke]{\pgfxy(-3,-2.6)}{\pgfpoint{2.2cm}{1cm}}
\pgfnoderect{data2}[stroke]{\pgfxy(3,-2.6)}{\pgfpoint{2.2cm}{1cm}}
\pgfputat{\pgfnodecenter{country}}{\pgfbox[center,center]{$\theta_{\textrm{country}}$}}
\pgfputat{\pgfnodecenter{urban}}{\pgfbox[center,center]{$E(\theta_{\textrm{urban}}) = \theta_{\textrm{country}}$}}
\pgfputat{\pgfnodecenter{rural}}{\pgfbox[center,center]{$E(\theta_{\textrm{rural}}) = \theta_{\textrm{country}}$}}
\pgfputat{\pgfnodecenter{data1}}{\pgfbox[center,center]{Urban Data}}
\pgfputat{\pgfnodecenter{data2}}{\pgfbox[center,center]{Rural Data}}
\pgfsetendarrow{\pgfarrowto}
\pgfnodeconnline{urban}{data1}    \pgfnodeconnline{rural}{data2}
\pgfnodeconnline{country}{urban}   \pgfnodeconnline{country}{rural}
\end{pgfpicture}
\end{center}
\vspace{15mm}
\caption{\label{fig-model}
A hierarchical model for input parameters of EPP.}
\end{figure}

Estimating parameters in dynamic models is often time consuming due to the lack of an analytic solution, the multi-modality and nonlinearity. Fitting the datasets from multiple areas simultaneously in the hierarchical model will further increase the computing cost. It is desirable to develop a procedure that produces the hierarchical model results without additional computational cost. Since running the dynamic model is the most time consuming part of the estimation, we propose applying the EPP model to fit each individual dataset as currently being implemented in EPP/Spectrum 2013 \shortcite{Brown2014} and using importance sampling to approximate the joint posterior distribution of parameters for multiple datasets. 

Note that the marginal distribution of $\theta_{ij}$ is $N(\mu_{0j}, \sigma_{0j}^2+\sigma_{1j}^2)$. We let $\mu_{0j}$ and $\sigma_{0j}^2+\sigma_{1j}^2$ be the mean and variance in equation (2) so that the marginal distribution of $\theta_{ij}$ is the same between the hierarchical model and the independent model (fitting the EPP model for each dataset independently). Let $\pi(\theta_{1.}, \ldots, \theta_{K.})$ be the joint prior density of $(\theta_{1.}, \ldots, \theta_{K.})$ for the independent model, which is simply the product of marginal densities of $\theta_{i.}$. Let $\pi_{\textup{HR}}(\theta_{1.}, \ldots, \theta_{K.})$ be the joint prior density of $(\theta_{1.}, \ldots, \theta_{K.})$ for the hierarchical model, and $\pi_{\textup{HR}}(\theta_{1.}, \ldots, \theta_{K.}) = \prod_j f_j(\theta_{1j}, \ldots, \theta_{Kj})$ for diagonal $\Sigma_0$ and $\Sigma_1$. After some derivation from equation (3), we can integrate out the national level parameter, $\mu$:
\begin{equation}
f_j(\theta_{1j}, \ldots, \theta_{Kj}) \propto \exp(-\frac{K \times \frac{\sum_{i=1}^K(\theta_{ij}-\bar{\theta}_{.j})^2}{\sigma_{1j}^2} + \frac{\sum_{i=1}^K(\theta_{ij}-\mu_{0j})^2}{\sigma_{0j}^2}}{2(K+ \frac{\sigma_{1j}^2}{\sigma_{0j}^2})}),
\end{equation}
where $\bar{\theta}_{.j} = \frac{\sum_{i=1}^K\theta_{ij}}{K}$. The parameter from the ith area will be favorable if it is close to the mean of the country level parameter, and to the mean of parameters from other areas. A small value of $\sigma_{1j}^2$ implies a high similarity of parameters across areas; and a small value of $\sigma_{0j}^2$ reflects a strong prior knowledge on the value of the country level parameter. Since $\sigma_{0j}^2+\sigma_{1j}^2$ is the variance of the original prior distribution of $\theta_{.j}$ in EPP/Spectrum, we only need to determine $\lambda_j = \sigma_{1j}^2/\sigma_{0j}^2$ which controls the fraction of variability that comes from the area effect. The smaller it is the more similar $\theta_{.j}$ is across areas. Here we use an empirical Bayes approach to determine the value of the hyper-parameter $\lambda$. We obtain a set of posterior medians by applying the independent model to 15 countries with high quality data (see result section for details), estimate between-country variance and the within-country variance for each parameter, and set the value of $\lambda_j$ as the ratio between corresponding variance estimates: $\lambda = (0.35,  0.24,  2.15,  0.28,  0.40,  2.19,  0.61,  0.12)$.

\subsection{Parameter Estimation via Importance Sampling}
Applying the independent model to each individual dataset, the posterior samples are drawn by the incremental mixture importance sampling (IMIS) algorithm which has been proved to be efficient for estimating parameters in dynamic systems with a moderate number of parameters \shortcite{Raftery2010,Brown2010}. It works as follows:
\begin{enumerate}
\item Draw initial samples from a sampling distribution, e.g. the prior distribution, $\pi(\theta_{i.})$;
\item Update the importance weight for each sample, which is the ratio between posterior density, $P(\theta_{i.}|\textrm{Data}_i)$, and density of the sampling distribution;
\item Find the sample with the highest weights, and draw new ones from a multivariate Gaussian distribution centered around it;
\item Combine new samples with all existing samples, update the sampling distribution by a mixture of initial sampling distribution and multiple Gaussian components, can recalculate the importance weight;
\item Iterate between step 2 and 4 until there is no large importance weight;
\item Resample all samples from the multinomial distribution with weights.
\end{enumerate}
The initial sampling distribution is often flat so that it ensures a good coverage of the entire parameter space. As a new Gaussian sampler is placed in the region that is under-represented by the sampling distribution at each iteration, the sampling distribution gets closer and closer to the posterior distribution.

To avoid a separate run of the hierarchical model, the step 6 of the IMIS algorithm can be revised as follows for drawing posterior samples from the hierarchical model:
\begin{enumerate}
\item[6.a.] Repeat steps $1\sim5$ for $i=1,\ldots,K$. All samples with non-ignoble importance weight will be stored, e.g. greater than $10^{-6}$,;
\item[6.b.] Create a joint sample of parameters across all areas by randomly taking one sample from each area, and repeat until we have a large number of candidate joint samples, e.g. 1,000,000; 
\item[6.c.] The importance weight of a joint sample $(\theta_{1.}, \ldots, \theta_{K.})$ is calculated as the product of the importance weights for $\theta_{i.}$'s times $\frac{\pi_{\textup{HR}}(\theta_{1.}, \ldots, \theta_{K.})}{\pi(\theta_{1.}, \ldots, \theta_{K.})}$, the ratio between the hierarchical model prior density and the independent model prior density;
\item[6.d.] Resample all samples from the multinomial distribution with weights.
\end{enumerate}


\subsection{Assessing Model Fit}
We compare performance of the independent model and the hierarchical model based on urban and rural datasets collected from the following 19 countries in Sub-Saharan Africa: Benin, Botswana, Burkina Faso, Burundi, Congo, Cameroun, Chad, Ethiopia, Ghana, Kenya, Lesotho, Mali, Central African Republic (RCA), Democratic Republic of the Congo (RDC), Rwanda, Tanzania, Uganda, Zambia and Zimbabwe. 

To assess the predictive ability of the model, a truncated version of each dataset is created by splitting the data years into three equal parts, removing the first and third part of ANC data, and keeping only the first NPBS data. The independent model is applied to each dataset and its truncated version. The hierarchical model results for areas of the same country are produced under the following two scenarios: (1) both areas with full data; (2) one area with truncated data and the other areas with full data. 

For Scenario 1, we calculate the expected log-likelihood by averaging the full data log-likelihood over all posterior samples, which is considered as a measure of the goodness of fit or the within-sample predictive accuracy \shortcite{Gelman2014}, the higher the value, the better the fit. For the truncated data in Scenario 2, we calculate the expected log-likelihood of the truncated data as well as the expected log-likelihood of the full data while only the truncated data is used for parameter estimation. The expected log-likelihood of the truncated data can be viewed as the goodness of fit when the data quality is low, and the expected log-likelihood of the full data measures the predictive accuracy for the whole epidemic time trend when the data is only available for a short period of time.

\section{Results}
\label{sect-Results}

We first focus on countries with relatively high-quality data in both urban and rural areas, so that the posterior distribution is more driven by the data instead of the prior distribution. Hereby, we define ``high-quality data" as a dataset with at least five years of antenatal clinic (ANC) data and one national population based survey (NPBS) data, and define ``data years" as the years for which we have data. This leads to 15 countries in Sub-Saharan Africa: Benin, Botswana, Burkina Faso, Burundi, Ethiopia, Ghana, Kenya, Lesotho, Mali,  Democratic Republic of the Congo, Rwanda, Tanzania, Uganda, Zambia,  Zimbabwe. 

We fit the r-trend model to each high-quality dataset and take the posterior medians as the point estimates. Table 1 summarizes the between-country and within-country standard deviation of those point estimates. It clearly shows that the variation of parameters across areas in the same country is smaller than the variation between countries, especially for $t_0$ -- starting year of the epidemic, $t_1$ -- the number of years that it takes to stabilize the epidemic, $\beta_0$ and $\beta_1$ describing the relationship between new infection rate and old infection rate, the clinic bias, $\beta_4$, which is the systematic difference between prevalence among pregnant women attending antenatal clinics and the general population. The ratios between within-variances and between-country variances provide empirical estimates $\hat{\lambda}$ for the hierarchical model.

\begin{table}[!h]
\caption{Between-country standard deviation $\sigma_0$, within-country standard deviation $\sigma_1$, and the ratio $\lambda=\sigma_1^2/\sigma_0^2$. Red color highlights the parameters whose within-country variation is much smaller than the between-country variation.}
\begin{tabular}{lcccccccc}
\hline
                  	& \textcolor{red}{$t_0$}   & \textcolor{red}{$t_1$} & $\log r_0$ & \textcolor{red}{$\beta_1$} & \textcolor{red}{$\beta_0$} & $\beta_2$ & $\beta_3$ & \textcolor{red}{$\textup{clinic bias}$}\\
\hline\hline
between  & 4.89     & 4.95       & 0.022     & 0.073  & 0.142     & 0.172   & 0.0037     & 0.110 \\
\hline
within      & 2.90     & 2.43       & 0.032     & 0.038  & 0.090     & 0.254   & 0.0029     & 0.037\\
\hline
$\lambda$      & 0.35     & 0.24       & 2.15     & 0.28  & 0.40     & 2.19   & 0.61     & 0.11\\
\hline
\end{tabular}
\end{table}

We then compare the performance between the hierarchical model and the independent model for all 19 countries, and summarize the improvement of expected log-likelihood from the independent model to the hierarchical model in Table 2. The difference in the expected log-likelihood between the hierarchical model and the independent model ranges from -3.8 to 5.4 in the (full, full) column in which the full data log-likelihood was calculated for the parameter values estimated from the full data; and ranges from -2.0 to 7.4 in the (trunc, trunc) column in which  the truncated data log-likelihood was calculated for the parameter values estimated from the truncated data. It suggests that the hierarchical model slightly improves the within-sample predictive accuracy, but the improvement is not substantial. The big difference between the independent model and the hierarchical model happens in the (full, trunc) column in which the full data log-likelihood was calculated for parameter values estimated from the truncated version of data. The difference ranges from -7.4 to 206.4. Among 38 datasets, there are 33 cases where the hierarchical model provides larger expected log-likelihood than the independent model, and the improvement is greater than 5 in 25 datasets and less than -5 in only one dataset. This suggests that the gain of expected log-likelihood in the hierarchical model mostly comes from its predictive ability.

\begin{table}[ht]
\caption{\footnotesize{A summary table for the number of data years, the number of antenatal clinics (ANC) and the improvement of expected log-likelihood: (full, full) means the full data log-likelihood where parameters are estimated from the full data; (full, trunc) means the full data log-likelihood where parameters are estimated from the truncated version of data; (trunc, trunc) means the truncated data log-likelihood where parameters are also estimated from the truncated data. Red color highlights the datasets with largest and smallest improvement for the (full, trunc) scenario.}}
\centering
\begin{tabular}{rlllll}
  \hline
 & Number of & Number of & \multicolumn{3}{c}{Improvement of expected log-likelihood} \\
 & data years & ANC sites & \footnotesize{(full, full)} & \footnotesize{(full, trunc)} & \footnotesize{(trunc, trunc)} \\ 
  \hline
  Benin Urban & 21 & 36 & 2.8 & 12.7 & 0.2 \\ 
  Benin Rural & 9 & 21 & 3.1 & 14.7 & 1.2 \\ 
  Botswana Urban & 17 & 10 & 2.2 & 136.8 & -0.2 \\ 
  Botswana Rural & 17 & 14 & 2.3 & 20.6 & 0.4 \\ 
  BurkinaFaso Urban & 20 & 7 & -0.2 & 14.1 & 0.2 \\ 
  BurkinaFaso Rural & 8 & 8 & 0.2 & \textcolor{red}{141} & -2 \\ 
  Burundi Urban & 20 & 7 & -0.9 & \textcolor{red}{-3.9} & 0.1 \\ 
  Burundi Rural & 20 & 9 & -3.8 & 3.8 & -0.5 \\ 
  Congo Urban & 8 & 10 & 0.2 & -2.6 & 0 \\ 
  Congo Rural & 5 & 10 & 0.7 & 4.3 & 0.1 \\ 
  Cameroun Urban & 15 & 35 & 3.4 & 36.4 & 0.3 \\ 
  Cameroun Rural & 3 & 40 & 5.1 & 29.2 & 1.9 \\ 
  Chad Urban & 8 & 18 & -0.8 & 3.9 & 0.1 \\ 
  Chad Rural & 3 & 6 & -0.3 & 0.4 & -0.2 \\ 
  Ethiopia Urban & 13 & 38 & 5.4 & \textcolor{red}{206.4} & 1.9 \\ 
  Ethiopia Rural & 9 & 73 & -2.8 & -1.3 & 1.9 \\ 
  Ghana Urban & 17 & 23 & 2 & 2.3 & 2.7 \\ 
  Ghana Rural & 16 & 17 & 1.8 & 15.3 & 2.3 \\ 
  Kenya Urban & 18 & 17 & 0.6 & 19.8 & -0.1 \\ 
  Kenya Rural & 15 & 23 & -0.8 & 23.7 & 0.6 \\ 
  Lesotho Urban & 10 & 2 & -0.7 & 4.6 & 0.1 \\ 
  Lesotho Rural & 10 & 11 & 2.1 & 48.2 & -0.1 \\ 
  Mali Urban & 5 & 13 & 1.7 & 34.7 & -0.2 \\ 
  Mali Rural & 5 & 7 & 2.6 & 54.7 & 0 \\ 
  RCA Urban & 14 & 29 & 0.4 & 5.4 & 0.2 \\ 
  RCA Rural & 3 & 9 & 1 & 1.1 & 0.7 \\ 
  RDC Urban & 23 & 20 & 4.2 & 62 & 0.5 \\ 
  RDC Rural & 14 & 27 & 0.7 & \textcolor{red}{148.7} & 7.4 \\ 
  Rwanda Urban & 8 & 21 & -1.3 & \textcolor{red}{-7.4} & -0.2 \\ 
  Rwanda Rural & 8 & 26 & -0.4 & -0.7 & 0.1 \\ 
  Tanzania Urban & 21 & 55 & 0.3 & 1.4 & 0.1 \\ 
  Tanzania Rural & 21 & 93 & 0.9 & 26.9 & 0.4 \\ 
  Uganda Urban & 22 & 14 & -1.6 & 12.2 & 0 \\ 
  Uganda Rural & 19 & 27 & -0.3 & 31.8 & 0 \\ 
  Zambia Urban & 7 & 12 & 1.2 & 119.1 & 0 \\ 
  Zambia Rural & 7 & 11 & 2.5 & \textcolor{red}{162.6} & 1.2 \\ 
  Zimbabwe Urban & 16 & 28 & -0.3 & 27.6 & 0 \\ 
  Zimbabwe Rural & 14 & 22 & 0.6 & 15.6 & 0.5 \\ 
   \hline
  Mean & 12.9 & 22.3 & 0.89 & 37.53 &  0.57 \\ 
  Standard Deviation & 6.1 & 18.6 & 1.97 & 54.08 & 1.42 \\ 
  \hline
\end{tabular}
\end{table}

In terms of the expected full data log-likelihood where parameters are estimated from the truncated version of the data, we highlight 4 datasets that have the largest improvement provided by the hierarchical model -- Burkina Faso rural, Ethiopia urban,  RDC rural, Zambia rural, and 2 datasets where the independent model works better than the hierarchical model -- Burundi rural and Rwanda urban. Figure 2 compares the estimated HIV prevalence trajectory between the independent model and the hierarchical model for those datasets. The hierarchical model results differ systematically from the independent model results in the top four sub-figures, and fit the observed ANC and NPBS data better, especially in the areas outside the red vertical lines which indicated the starting and end year of truncated data. The independent model and the hierarchical model produce similar prevalence trends in the bottom two sub-figures, and their difference in expected log-likelihood is not substantial.

\begin{figure}[h]
\begin{tabular}{cc}
\begin{minipage}{6.5cm}
\includegraphics[width=6.5cm]{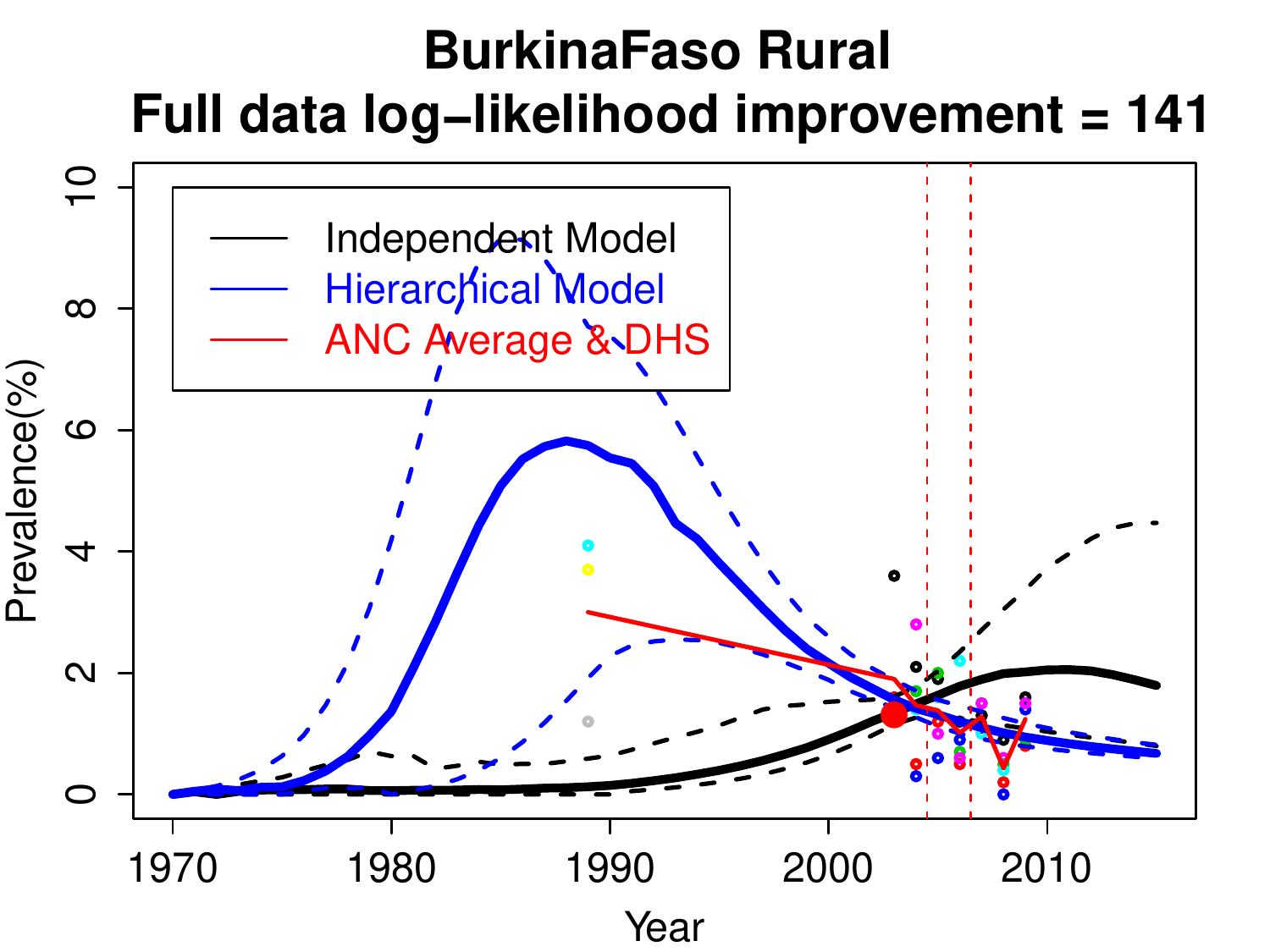}
\end{minipage}
&
\begin{minipage}{6.5cm}
\includegraphics[width=6.5cm]{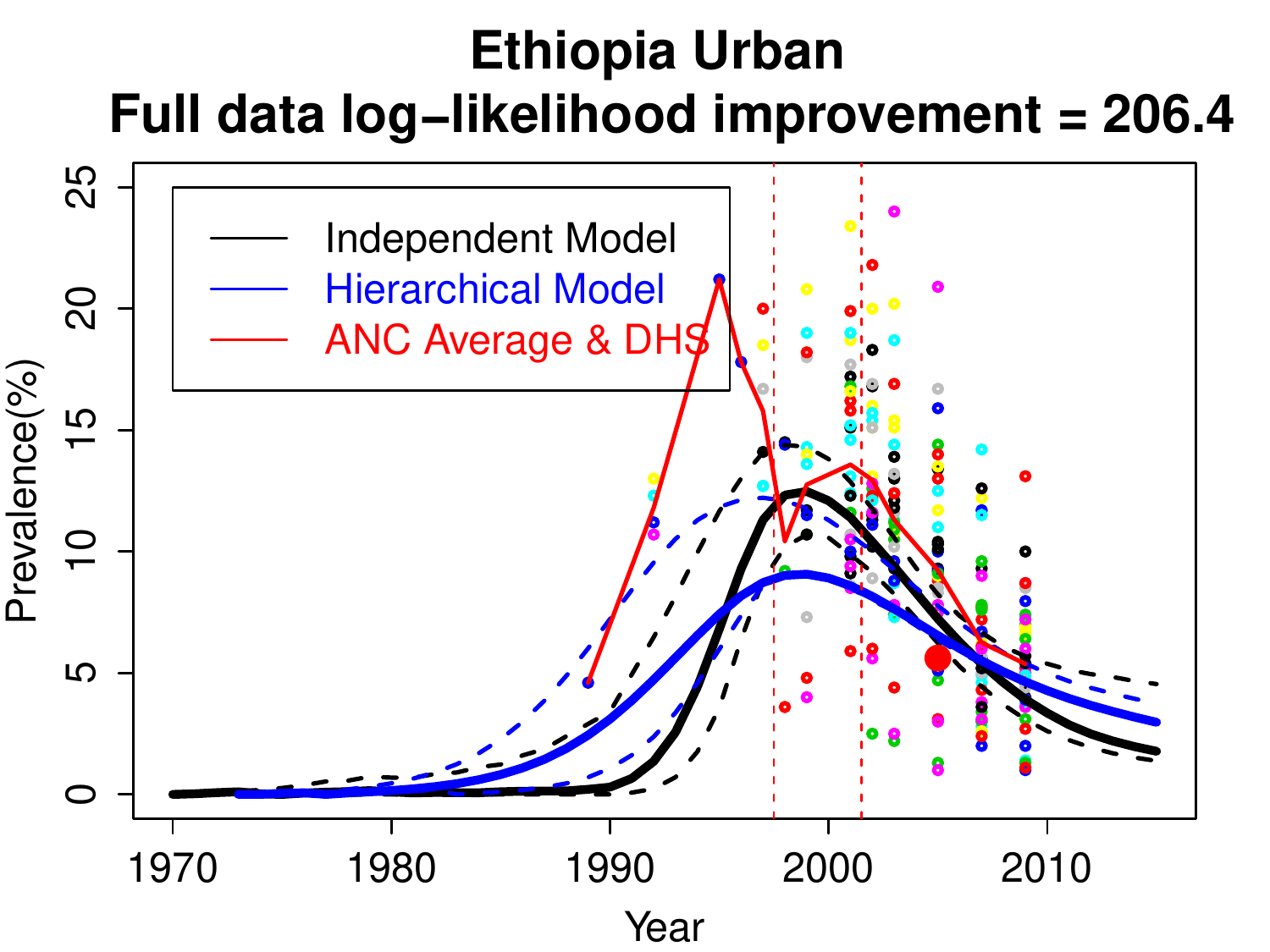}
\end{minipage}
\\
\begin{minipage}{6.5cm}
\includegraphics[width=6.5cm]{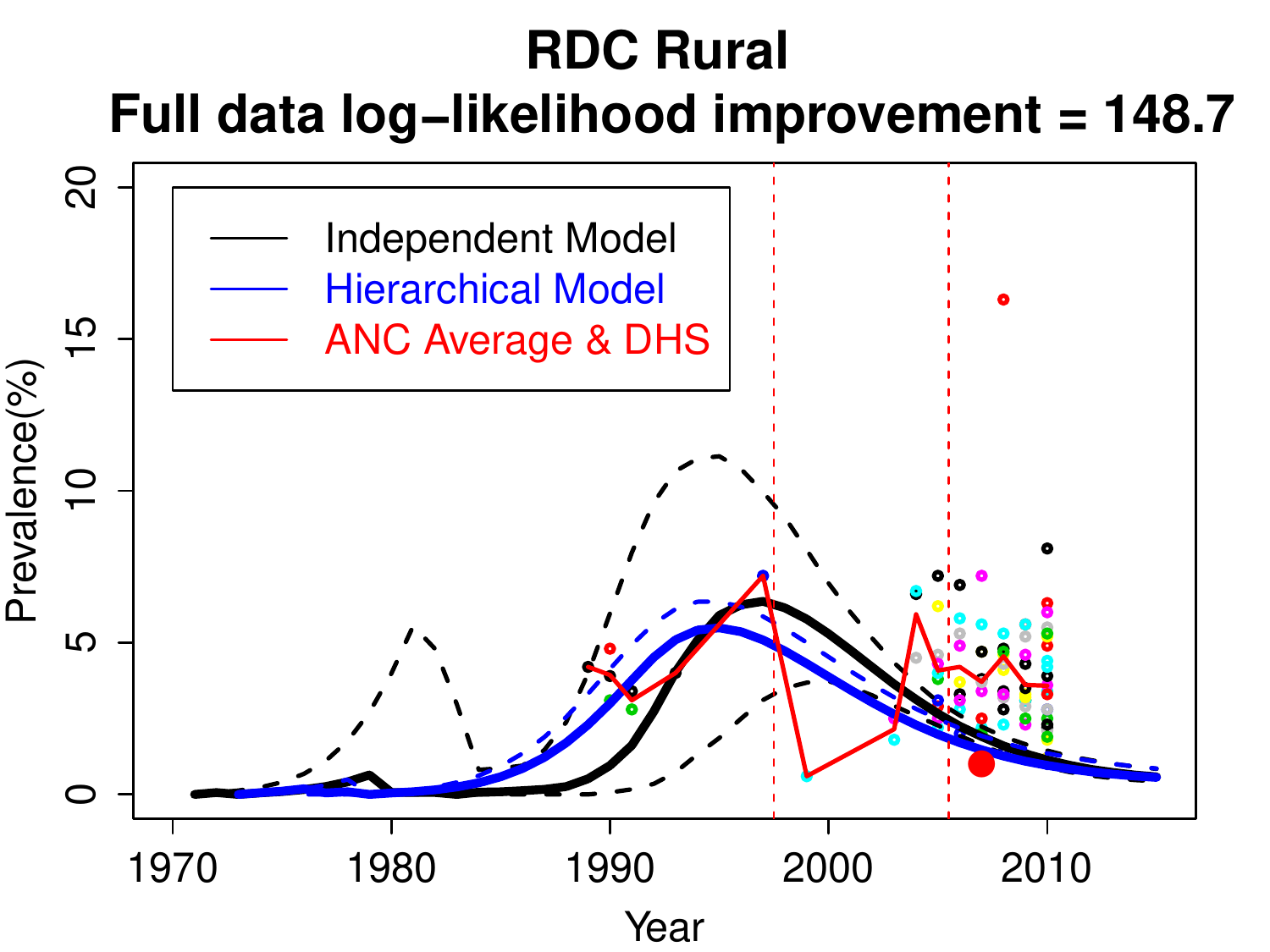}
\end{minipage}
&
\begin{minipage}{6.5cm}
\includegraphics[width=6.5cm]{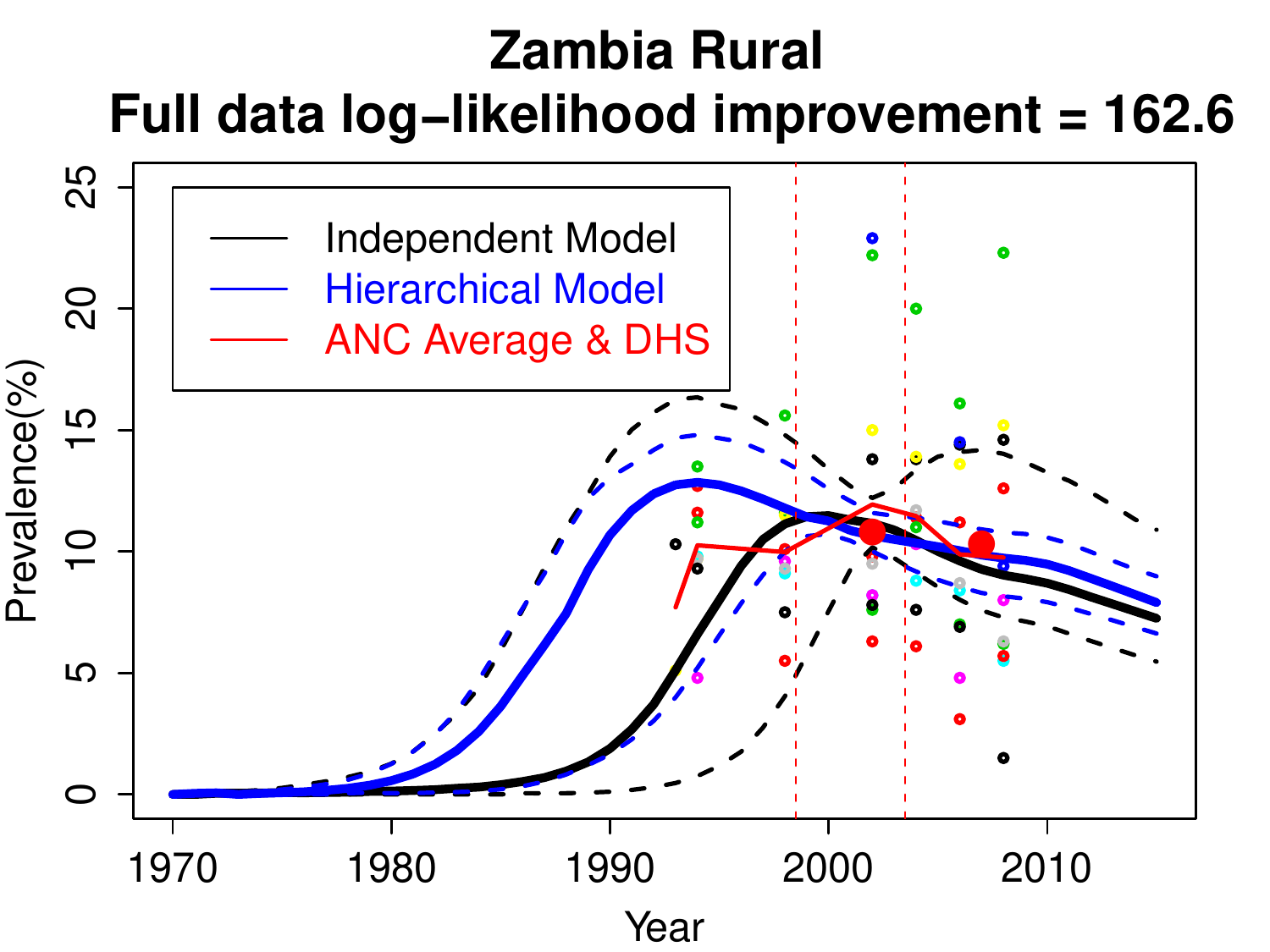}
\end{minipage}
\\
\begin{minipage}{6.5cm}
\includegraphics[width=6.5cm]{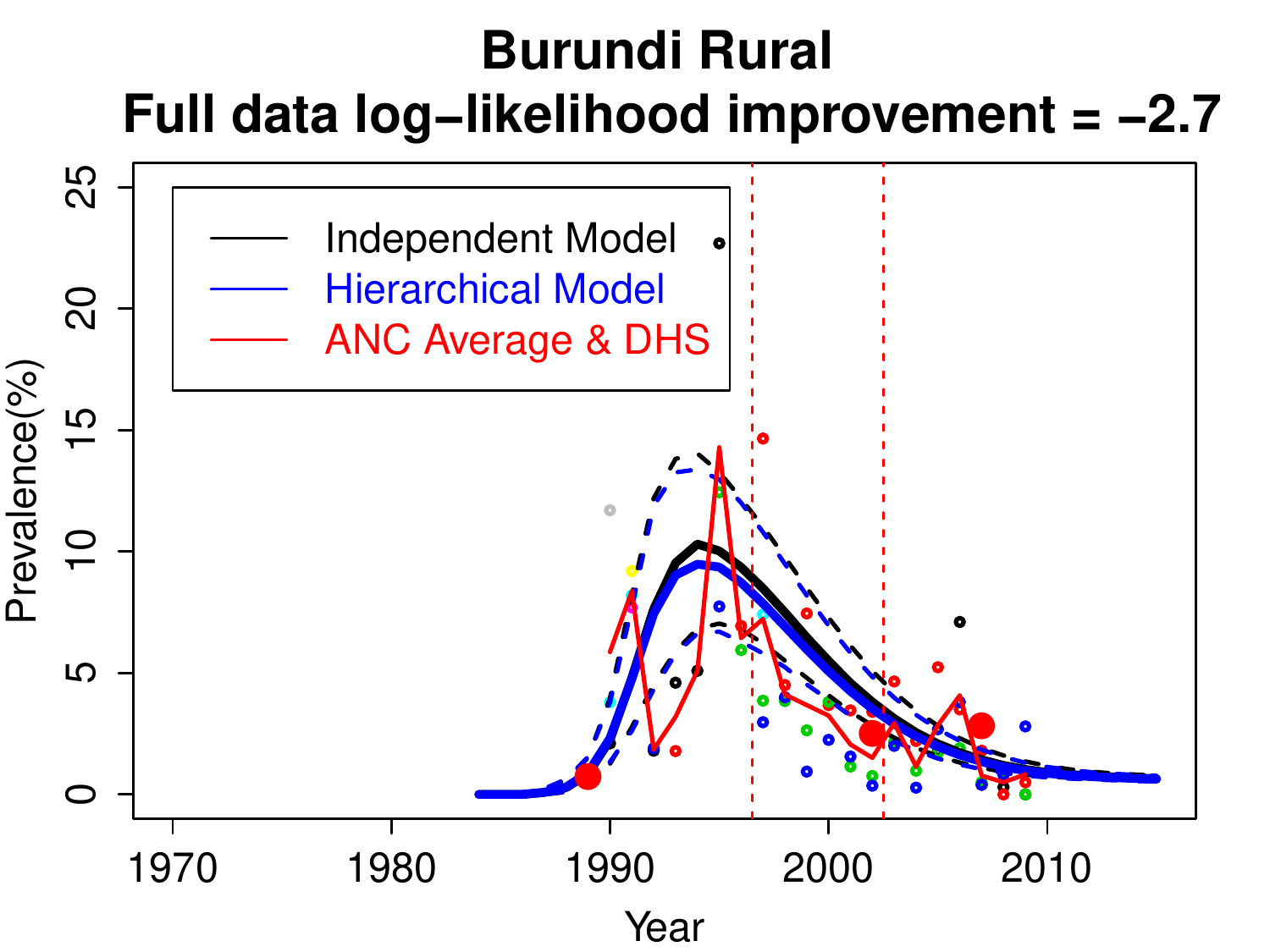}
\end{minipage}
&
\begin{minipage}{6.5cm}
\includegraphics[width=6.5cm]{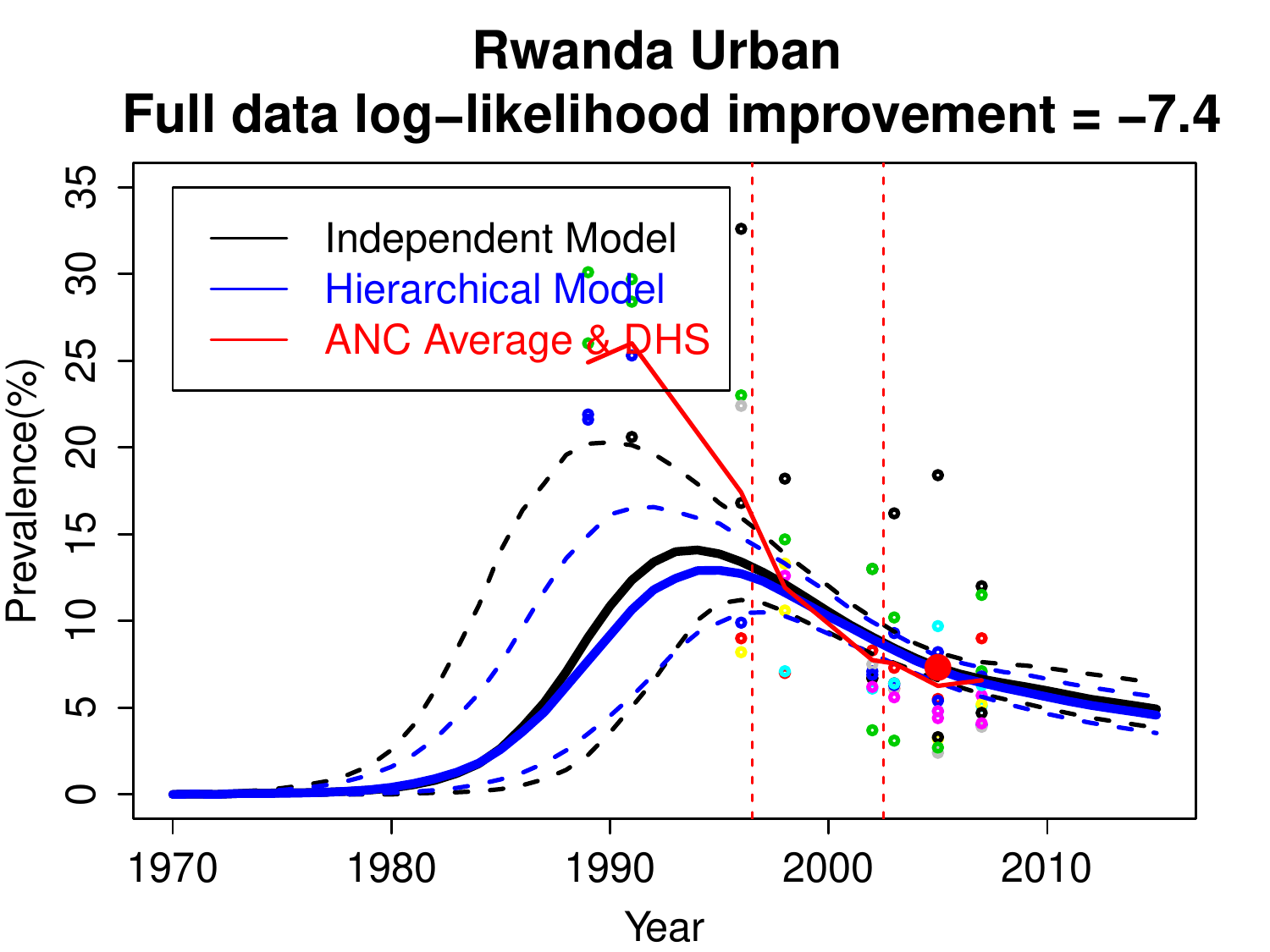}
\end{minipage}
\\
\end{tabular}
\caption{\footnotesize{HIV prevalence estimation for 4 datasets that have the largest improvement provided by the hierarchical model, and 2 datasets where the independent model works better than the hierarchical model. Within each figure, the black solid line is the posterior median trajectory of the independent model; the  blue solid line is the posterior median trajectory of the hierarchical model; the dashed black/blue lines show the 90\% credible intervals; colored dots are observed prevalence from different ANC sites; the red solid line is the raw average trajectory of ANC prevalence; the large red dots are prevalence estimated in a national population based survey (NPBS). The red dashed vertical lines show the window of truncated data.}}
\end{figure}

Nigeria is one of the countries that plans to estimate and predict HIV impact at the state level, and it comprises 36 states and the Federal Capital Territory, Abuja. The data quality varies across states: the number of years for which ANC data is available ranges from 6 to 9; the number of clinic sites ranges from 2 to 8. For each of the 37 areas, we estimate model parameters by using the truncated dataset with the combination of full datasets in other areas, and then calculate the full data log-likelihood. The difference of the expected log-likelihood of the full data between the hierarchical model and the independent model ranges from -6.7 to 78.5 with mean improvement 12.3, and we get positive improvement in 34 states. Figure 3 compares the estimated HIV prevalence trajectory between the independent model and the hierarchical model for 6 states where the absolute difference of the expected log-likelihood of the full data between the hierarchical model and the independent model is greater than 20.


\begin{figure}[h]
\begin{tabular}{cc}
\begin{minipage}{6.5cm}
\includegraphics[width=6.5cm]{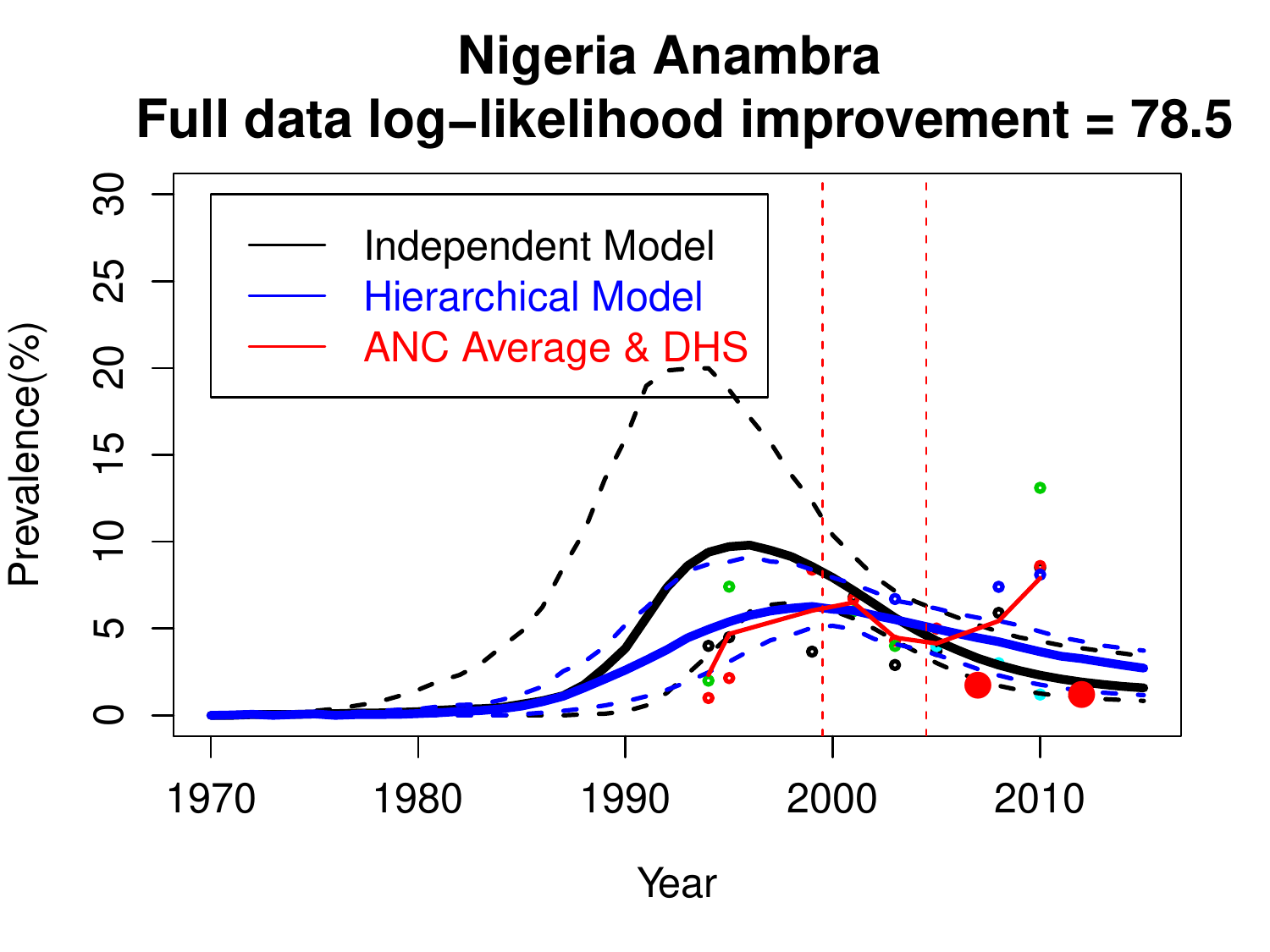}
\end{minipage}
&
\begin{minipage}{6.5cm}
\includegraphics[width=6.5cm]{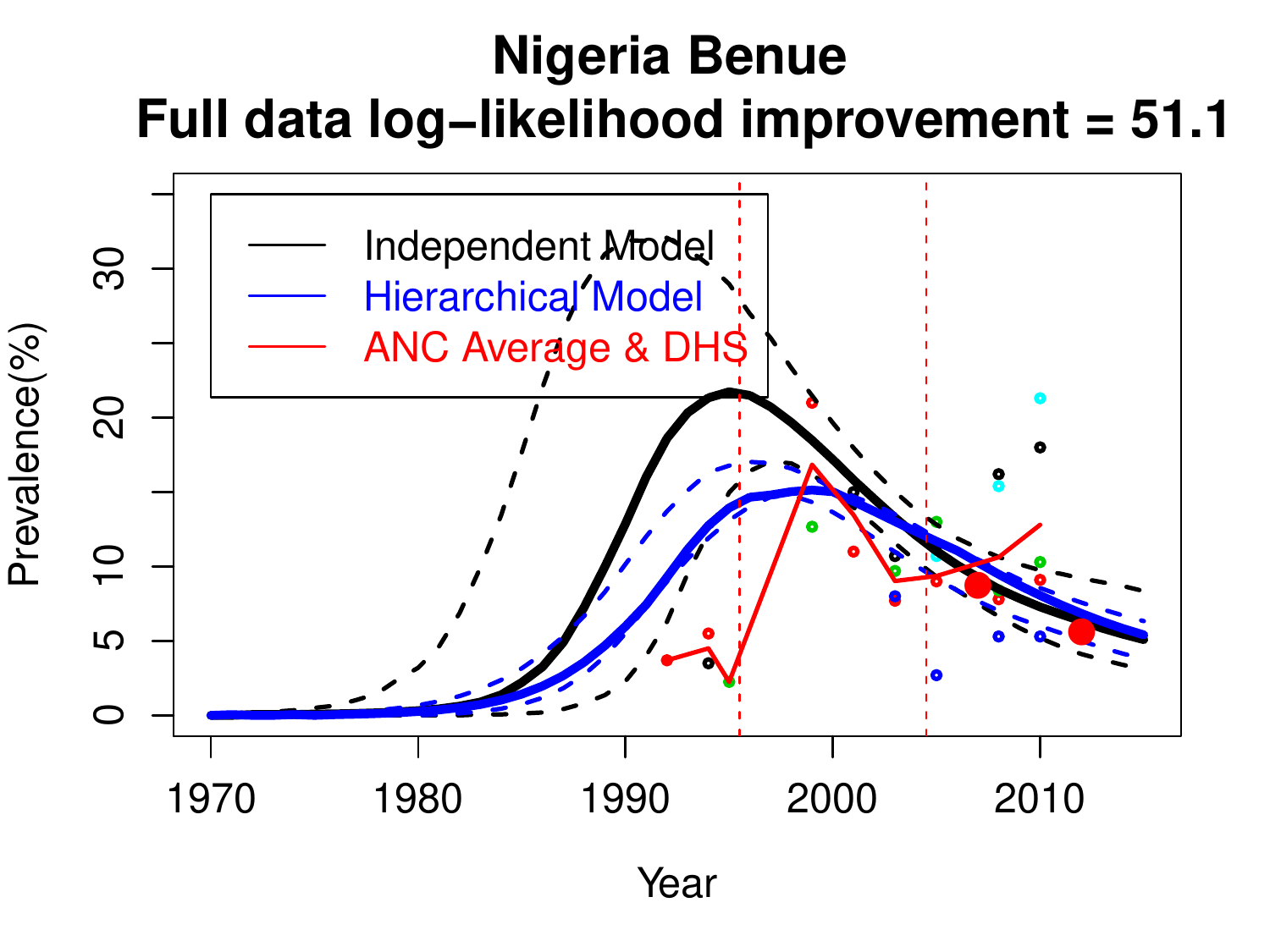}
\end{minipage}
\\
\begin{minipage}{6.5cm}
\includegraphics[width=6.5cm]{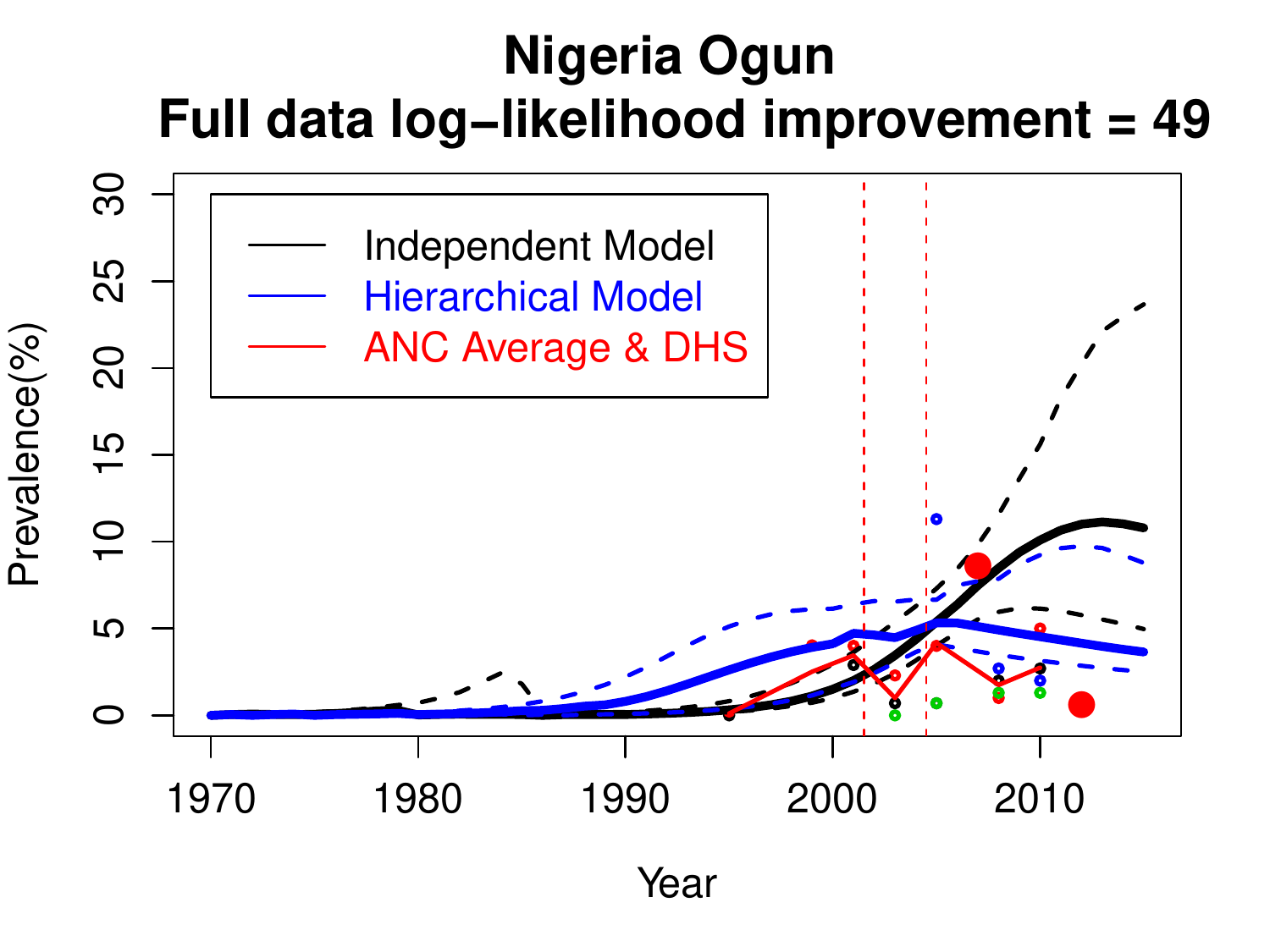}
\end{minipage}
&
\begin{minipage}{6.5cm}
\includegraphics[width=6.5cm]{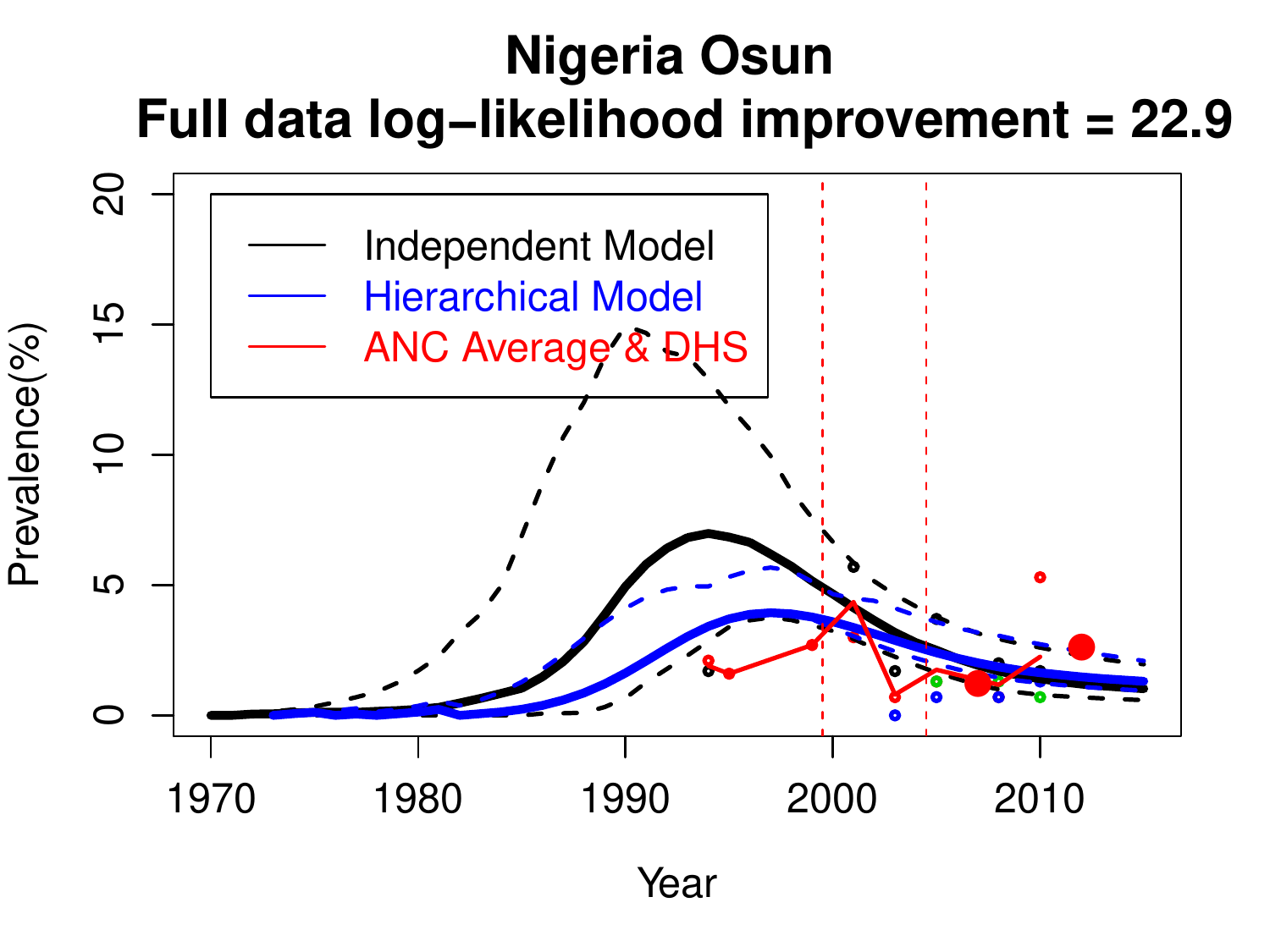}
\end{minipage}
\\
\end{tabular}
\caption{\footnotesize{HIV prevalence estimation for 4 states in Nigeria where the improvement of the expected log-likelihood of the full data provided by the hierarchical model is greater than 20. Within each figure, the black solid line is the posterior median trajectory of the independent model; the  blue solid line is the posterior median trajectory of the hierarchical model; the dashed black/blue lines show the 90\% credible intervals; colored dots are observed prevalence from different ANC sites; the red solid line is the raw average trajectory of ANC prevalence; the large red dots are prevalence estimated in a national population based survey (NPBS). The red dashed vertical lines show the window of truncated data.}}
\end{figure}

In addition to the examination of the marginal distribution of prevalence within each state, the estimated prevalence levels in states are correlated in the hierarchical model, but not so in the independent model. Therefore, the hierarchical model provided a more realistic joint distribution of the quantity of interest across areas, e.g. prevalence, incidence, or mortality. We apply the hierarchical model to full datasets in 36 states and the capital of Nigeria, and calculate the correlation between any pair of states at each year. Figure 4 shows the examples of prevalence correlations and incidence correlations between federal capital territory, Abuja, and its 3 neighboring states. Since we have assumed a hierarchical structure for the EPP input parameters which summarize the characteristics of the HIV epidemic, the correlations of the EPP output prevalence and incidence vary across time. Those correlations are mostly positive due to the positive correlation of EPP input parameters, and can be as high as 0.6 for some pairs of states. It is relatively lower in years where the data is rich within the sub-national area so that there is no need to borrow strength from other areas. 

\begin{figure}[h]
\includegraphics[width=13cm]{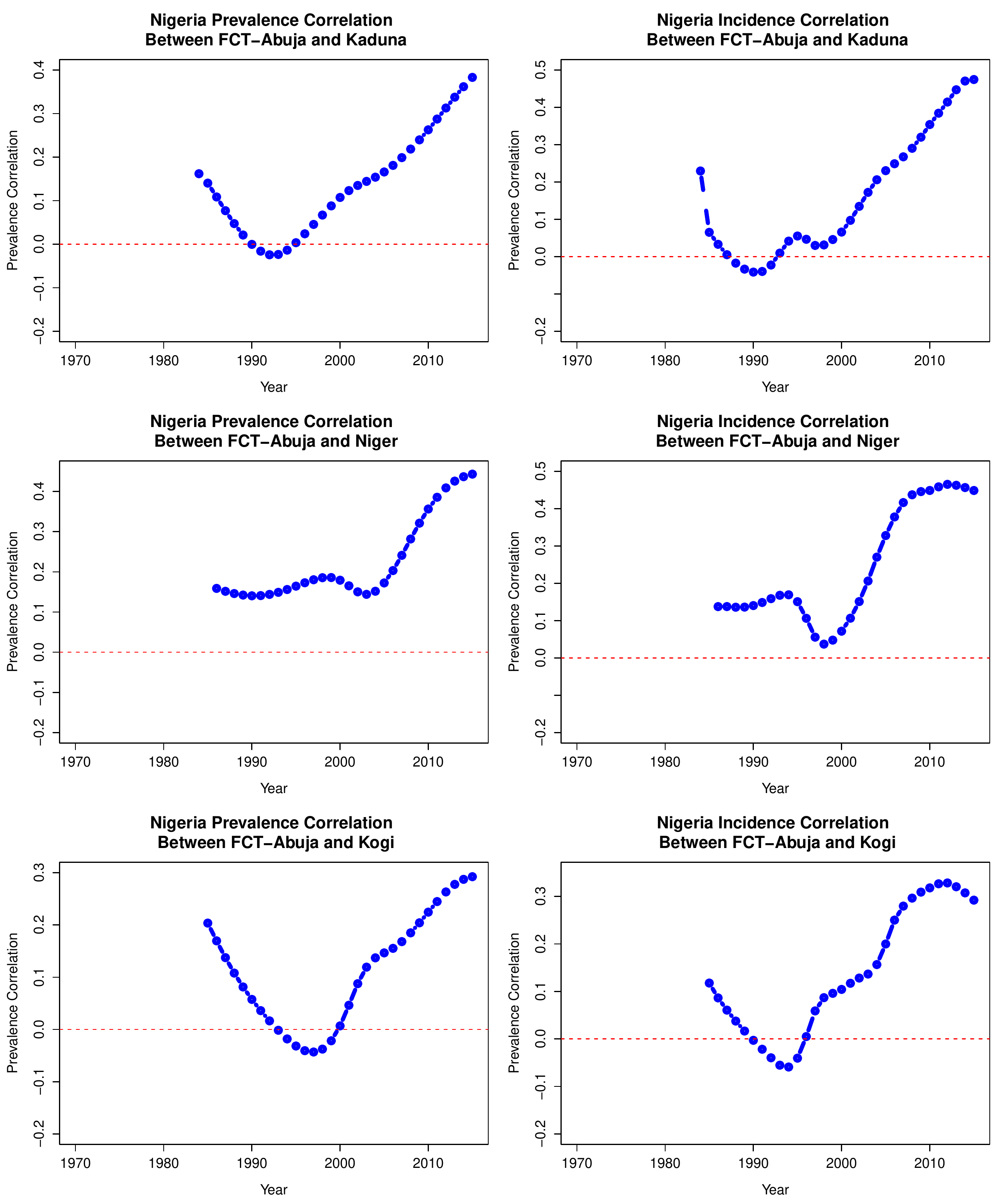}
\caption{\footnotesize{The correlations of prevalence estimates between federal capital territory, Abuja, and its 3 neighboring states at each year: prevalence on the left and incidence on the right. The blue lines are correlation in the hierarchical model, and the red dashed line indicates a zero correlation which corresponds to the expected correlation in the independent model.}}
\end{figure}

\section{Discussion}
\label{sect-Discussion}
Planning an effective response to the HIV epidemic and assessing the impact of the past response requires quantitative analysis. Sub-national estimates can be used for better program planning and management, such as assessing and meeting the needs for commodities, human resources and other program elements, measuring population coverage of treatments, and monitoring and evaluating interventions. In this article, we describe a hierarchical epidemiological method that assumes the patterns of the sub-national epidemic are similar. For the estimation and projection of HIV epidemics, this will be the first attempt of developing a joint model for multiple areas within a country. It will be implemented in the next round of EPP/Spectrum software (available from http://www.futuresinstitute.org). The proposed model can be easily generalized to the study of other epidemics. 

The computing cost of estimating parameters in a dynamic model is often high due to the lack of an analytic solution, the multi-modality and nonlinearity. We propose an importance sampling method that draws posterior samples of parameters in the hierarchical model without needing to refit the data or unpack the existing software for implementing an independent model. We first obtain posterior samples of parameters for each datasets independently. Since the hierarchical model and the independent model only differ at their prior distributions, we can randomly combine posterior samples of the independent model from multiple areas, reweight the joint samples by the ratio between the independent model prior and the hierarchical model prior, and draw a set of joint posterior samples from the multinomial distribution with weights. Note that the joint sample with a large weight can result in a large number of replicates in the posterior samples of parameters in the hierarchical model. We may need a large number of random combinations to ensure a sufficient number of unique posterior samples. A more efficient sampling strategy should be developed in the future.

When a country has a large geographical territory or diverse regions of public health conditions, it is worthy to further develop a spatial model or multilevel hierarchical model based on the knowledge of the national officials. In those cases, the similarity of epidemic patterns may depend on the distance or adjacency status between areas, or the regions to which the areas belong. In countries with low-level and concentrated epidemics, HIV has spread rapidly in several high risk groups, but is not well established in the general population. Fewer data are available for those high risk populations due to the stigmatized nature of these populations in many countries. In such cases, a hierarchical model can allow sharing information across areas and high risk groups.

Note that, in this specific application, the incremental mixture importance sampling method is used to draw posterior samples for each individual data, but our proposed reweighting approach is generic and does not depend on how the posterior samples are drawn for individual datasets. For instance, if Markov chain Monte Carlo (MCMC) is used to draw posterior samples for individual datasets with reasonable burn-in and thinning, we can assume that those posterior samples have equal weights in the independent model. This suggests an alternative way of estimating parameters in dependent data, e.g. data with hierarchical/spatial/temporal structure: fitting each piece of data independently, and merge the results with adjustments for the difference between the independent model and the joint model. 

The results presented in this paper are based on illustrative HIV prevalence data for these countries, which may not be complete. These results should therefore not be seen as replacing or competing with official estimates regularly published by countries and UNAIDS.

\bibliographystyle{imsart-nameyear}
\bibliography{EPP}


\end{document}